

\input phyzzx

\input epsf

\newif\iffigsinc
\ifx\epsfbox\UnDeFiNeD\message{(NO epsf.tex, FIGURES WILL BE IGNORED)}
\figsincfalse
\else\message{(FIGURES WILL BE INCLUDED)}\figsinctrue\fi
\def\hr{{\bf \hat r}}
\def\c{{\bf C}}
\def\cl{{\bf C}^{(\lambda)}}
\def\cls{{\bf C}^{(\lambda)*}}
\def\clp{{\bf C}^{(\lambda')}}
\def\J{{\cal J}}

\def\splus{k_+}
\def\sminus{k_-}
\def\high{\vphantom{\Biggl(}\displaystyle}

\def\pr#1#2#3#4{Phys. Rev. D{\bf #1}, #2 (19#3#4)}
\def\np#1#2#3#4{Nucl. Phys. {\bf B#1}, #2 (19#3#4)}

\def\prl#1#2#3#4{Phys. Rev. Lett. {\bf #1}, #2 (19#3#4)}

\def\rn{Reissner-Nordstr\"om }

\nopagenumbers
\line{\hfil CU-TP-651}
\line{\hfil hep-th/9409013 }
\vglue .5in
\centerline{\twelvebf Instabilities of magnetically charged black
holes}
\vskip .3in
\centerline{\it S.~Alexander Ridgway and Erick J.~Weinberg}
\vskip .1in
\centerline{Physics Department, Columbia University}
\centerline{New York, New York 10027}
\vskip .4in
\baselineskip=14pt
\overfullrule=0pt
\centerline {\bf Abstract}

The stability of the magnetically charged Reissner-Nordstr\"om black hole
solution is investigated in the context of a theory with massive
charged vector mesons.  By exploiting the spherical symmetry of the
problem, the linear perturbations about the Reissner-Nordstr\"om solution
can be decomposed into modes of definite angular momentum $J$.  For
each value of $J$, unstable modes appear if the horizon radius is less
than a critical value that depends on the vector meson gyromagnetic
ratio $g$ and the monopole magnetic charge $q/e$.  It is shown that
such a critical radius exists (except in the anomalous case $q={1\over
2}$ with $0 \le g \le 2$), provided only that the vector meson mass is
not too close to the Planck mass.  The value of the critical radius is
determined numerically for a number of values of $J$.  The
instabilities found here imply the existence of stable solutions with
nonzero vector fields (``hair'') outside the horizon; unless $q=1$ and
$g>0$, these will not be spherically symmetric.

\vskip .1in
\noindent\footnote{}{\twelvepoint \noindent This work was supported in part by
the US Department of Energy.}

\vfill\eject

\baselineskip=20pt
\pagenumbers
\pageno=1

\chapter{Introduction}

    The Reissner-Nordstr\"om solution of the coupled Maxwell-Einstein
equations describes a spherically symmetric black hole endowed with
electric or magnetic charge.  It may remain a solution when the Maxwell
theory is embedded in a larger theory. In particular, a trivial
extension of the Reissner-Nordstr\"om solution gives a magnetically
charged solution to a spontaneously broken $SU(2)$ gauge theory.
However, it has been shown\Ref\lnw{ K. Lee, V.P. Nair and E.J. Weinberg, Phys.
Rev. Lett. {\bf 68}, 1100 (1992).}
that if the black hole horizon is
sufficiently small (roughly, less than the Compton wavelength of the
massive vector meson) this solution develops a classical instability.
This instability suggests that there should be some other, stable
solution with the same mass and magnetic charge.  This solution has
been found\Ref\bhmono{K.~Lee, V.P.~Nair and E.J.~Weinberg,
\pr{45}{2751}92.};
it is a gravitationally perturbed magnetic monopole with a
black hole inside it.

The spontaneously broken gauge theory can be viewed as a special case of a
much wider class of theories obtained by varying the vector meson
magnetic moment and by allowing a less restrictive set of
interactions. Although such theories will not in general be
renormalizable, they are perfectly sensible classically.   The
magnetically charged solutions of these theories
have been studied\Ref\leew{K.~Lee and
E.J.~Weinberg, \prl{73}{1203}94. }
both with and without gravity.  Without gravity, the solutions are
singular unless certain relations between the couplings are satisfied.
When gravity is added, there are both Reissner-Nordstr\"om-type
solutions and, for
certain ranges of parameters, magnetically charged solutions with
nontrivial matter fields outside the horizon.  In this paper we
analyze the stability of the \rn solutions against small
perturbations.  We determine the critical value of the horizon radius
below which the \rn solution is unstable, and for which the less
trivial new solutions are guaranteed to exist. In a future paper we will show
how the results of the present analysis can be used to construct such
new solutions.

    The Reissner-Nordstr\"om metric may be written in the form
$$  ds^2 = -B(r) dt^2 + B^{-1}(r) dr^2
    + r^2 d\theta^2 + r^2\sin^2\theta d\phi^2 ,
      \eqno\eq$$
where
$$ \eqalign{  B(r)
     &= 1 - {2MG \over r} + {4\pi G (Q_M^2 + Q_E^2) \over r^2} \cr
      &= \left( 1 -{r_+\over r} \right) \left( 1 -{r_-\over r} \right) ,}
       \eqno\eq$$
with $Q_M$ and $Q_E$ being the magnetic and electric charges,
respectively.  We consider here the purely magnetic case, with $Q_E=0$
and $Q_M=q/e$, where the Dirac quantization condition restricts $q$ to
be either an integer or a half-integer.  Without loss of generality,
we may assume that $q$ is positive.  In order that the singularity at
$r=0$ be hidden within a horizon, we assume that $M$ is greater than
or equal to the extremal mass
$M_{\rm ext}=\sqrt{4\pi} (q/e) M_{\rm Pl}$, and hence that the outer
horizon $r_+ \equiv r_H \ge  \sqrt{4 \pi} (q/e) M_{\rm Pl}^{-1}$.
The  magnetic charge gives rise to a radial magnetic field
$$ F_{\theta\phi} = {q \over e} \, \sin\theta .
     \eqno\eq$$
All other components of $F_{\mu\nu}$ vanish.

     We want to consider the effect on this solution of
introducing a massive charged vector field $W_\mu$.  This field might
be a non-Abelian gauge field which has acquired a mass through the
Higgs mechanism (this was the only case considered in Ref.~[\lnw]), but
this need not be the case.  For our purposes here, we will
only need the terms in the
action which are quadratic in $W$.  These take the form
$$\eqalign{ S_{\rm quad} = \int d^4x \sqrt{- \det(g_{\mu\nu})} \Biggl[
        & -{1\over 2} \left| D_\mu W_\nu - D_\nu W_\mu \right|^2
        - m^2 \left| W_\mu \right|^2 \qquad\cr
   &\qquad -{ieg \over 4} F^{\mu\nu} \left( W^*_\mu W_\nu - W^*_\nu W_\mu
    \right) \Biggr] . }
\eqno\eq $$
Here $ F_{\mu\nu} = \partial_\mu A_\nu -\partial_\nu A_\mu$ is the
electromagnetic field strength while $D_\mu = \partial_\mu - ie A_\mu$
is the gauge covariant derivative.  The $W$ mass, $m$, might depend on
the value of some scalar field $\phi$, as in the Higgs case, but this
dependence can be ignored here; $m$ is then the value corresponding to
the vacuum value of $\phi$.  The last term shown is a magnetic moment
term, with the gyromagnetic factor $g$ arbitrary.  The integration is
over the region of spacetime outside the horizon of the black hole; it
is this region that is relevant for investigation of black hole
stability.  We must therefore impose boundary conditions at the horizon,
which will be discussed later.

     The Reissner-Nordstr\"om solution is trivially extended to this
theory by setting $W_\mu=0$ and setting all scalar fields at their
vacuum values everywhere.  We now want to consider fluctuations about
this solution.  These can be divided into several decoupled sets,
comprising the metric and electromagnetic perturbations, the scalar
field perturbations, and the vector field perturbations.
The first set has been analyzed in detail\ref{V.~Moncrief, \pr{9}{2707}74;
V.~Moncrief, \pr{10}{1057}74; V.~Moncrief, \pr{12}{1526}75.}
and shown not to lead to any
instability.  Because the scalar fields are assumed to be at their
vacuum values, it is easy to show that small perturbations in these
fields also cannot give rise to any instability; the proof parallels the
proof\ref{J.D.~Bekenstein, \pr{5}{1239}72.}
of the no-hair theorem for a free massive scalar field.
The analysis of the
remaining perturbations, those in $W_\mu$, is the subject of this
paper.

     The remainder of this paper is organized as follows.  In Sec.~2,
we exploit the spherical symmetry of the unperturbed solution
by expanding $W_\mu$ as a sum of eigenmodes of angular momentum
and then decomposing the quadratic action into a sum of terms, each of
which involves only modes of a given angular momentum.  In Sec.~3, we
show how both $W_t$ and $W_r$ can be eliminated, thus reducing the
problem to one involving at most two radial functions for each value of
the angular momentum.  In Sec.~4 we show that instabilities develop if
the horizon radius $r_H$ is less than some critical value $r_{\rm
cr}$, and obtain some analytic bounds on this critical value.
Section~5 describes our numerical results for this critical radius.
Some concluding remarks are contained in Sec.~6.

\chapter{Angular momentum decomposition of modes}

The spherical symmetry of the underlying solution allows
the perturbations to be decomposed into modes of definite angular
momentum.  Because of the nontrivial transformation properties of the
monopole vector potential under rotations (or, more fundamentally,
because of the extra angular momentum arising from any
charge-monopole pair) one cannot use the ordinary spherical harmonics
for this decomposition, but must instead employ monopole
spherical harmonics.  We begin this section by reviewing the
properties of these harmonics.  Although our eventual interest is in a
curved space-time, for the sake of compactness we adopt here a flat space
notation with vectors denoted by bold face.

For a scalar field of charge $e$, the appropriate spherical
harmonics are the Wu-Yang monopole harmonics
$Y_{qLM}(\theta, \phi)$\ref{I.~Tamm, Z.~Phys.~{\bf 71}, 141 (1931);\hfil\break
T.T.~Wu and C.N.~Yang, \np{107}{365}76.}, which
are eigenfunctions of the orbital angular momentum
$$  {\bf L} = -i{\bf r}\times {\bf D} - q {\bf \hr}
        \eqno\eq $$
obeying
$$ \eqalign{ {\bf L}^2 Y_{qLM} &= L(L+1) Y_{qLM} \cr
       L_z Y_{qLM} &= M Y_{qLM}  .}
     \eqn\scaleq $$
The quantum numbers $L$ and $M$ take on the values
$$  \eqalign{  L &= q, \, q+1, \,\dots \cr
               M& = -L,\, -L+1,\, \dots ,L .}
     \eqno\eq $$
These harmonics form a complete orthonormal set, with
$$ \int d\Omega\, Y_{qLM}^* Y_{qL'M'} \equiv \int_0^\pi \sin\theta \, d\theta
   \int_0^{2\pi}d \phi \,  Y_{qLM}^* Y_{qL'M'} = \delta_{LL'}\,\delta_{MM'} .
     \eqn\scalarnorm$$

  For spin-one fields one must use monopole vector harmonics, which
are eigenfunctions of ${\bf J} = {\bf L} + {\bf S}$.  In general,
there will be several of these corresponding to the same values of
${\bf J}^2$ and $J_z$.  We will find it convenient to
use a set\ref{E.J.~Weinberg, \pr{49}{1086}94. },
denoted by ${\bf C}^{(\lambda)}_{qJM}$, in which these are
distinguished by the value of $\bf \hat r\cdot S$.
These obey
$$\eqalign{ {\bf J}^2\, \cl_{qJM} &=  J(J+1) \,\cl_{qJM} \cr
    J_z \,\cl_{qJM} &= M \,\cl_{qJM} \cr
    {\bf \hr \cdot S} \,\cl_{qJM} &= i{\bf \hr}\times \cl_{qJM}
     =\lambda \,\cl_{qJM} }
     \eqn\eigenvalues $$
and are normalized so that
$$ \int d\Omega\, \cls_{qJM} \cdot \clp_{qJ'M'}
      = {\delta_{JJ'}\,\delta_{MM'}\,\delta_{\lambda\lambda'} \over r^2} .
           \eqn\vectornorm $$

    The allowed values for the total angular momentum quantum
number $J$ occur in integer steps beginning with $q-1$,
unless $q=0$ or $1/2$, in which case the minimum value is $J=q$. As
usual, the values of $M$ run from $-J$ to $J$ in integer steps.
Generically, there are three sets of harmonics for each $J$, with
$\lambda = 1$, 0,  and $-1$.  However, some of these are absent if $J$
is equal to $q$ or $q-1$.  For $J=q-1$ there is only a single family
of harmonics, with $\lambda=1$; these harmonics, which will be of
particular importance in our analysis, have the important property
that both their covariant curl and their covariant divergence vanish.
For $J=q>0$ there are two families of harmonics, with  $\lambda = 1$
and 0.  Finally, for $J=q=0$ (i.e., ordinary vector spherical
harmonics), only $\lambda=0$ is present.

    The last line of Eq.~\eigenvalues\ shows that the $\lambda=0$
harmonics must be purely radial; the normalization condition
\vectornorm\ then fixes them (up to an arbitrary phase) to be
$$ {\bf C}^{(0)}_{qJM} = {1 \over r}{\bf \hat r}Y_{qJM} .
   \eqn\zero $$
Since, as one can easily show,
vector harmonics with different values of $\lambda$ are orthogonal at
each point in space, those with $\lambda = \pm 1$ can only have angular
components.

     We will need the formulas for the covariant curls of the vector
harmonics.  For writing these, it is useful to define
$$   \J^2 = J(J+1) -q^2
\eqno\eq$$
and
$$ k_\pm = \sqrt{{\cal J}^2 \pm q \over 2} .
\eqno\eq $$
The curls can then be written as
$$ \eqalign{ {\bf D} \times {\bf C}^{(\pm 1)}_{qJM}
     &= \pm  {i k_\pm \over r} {\bf C}^{(0)}_{qJM}  \cr
   {\bf D} \times {\bf C}^{(0)}_{qJM}
    &= \pm  {i \over r} \left[ k_+ {\bf C}^{(1)}_{qJM}
           - k_- {\bf C}^{(-1)}_{qJM} \right] . }
\eqno\eq$$
(Note that $k_+=0$ if $J=q-1$, implying that the corresponding
harmonics are indeed curl-free, as asserted above.)

We will also make use of the identity
$$  {\bf D} Y_{qJM} =  k_+ {\bf C}^{(1)}_{qJM}
       + k_- {\bf C}^{(-1)}_{qJM} .
     \eqno\eq$$

   Let us now proceed to the decomposition of the quadratic action.
We begin with a mode expansion of $W_\mu$.  Recalling that the
$\lambda=0$ harmonics are purely radial while those with $\lambda =
\pm 1$ are purely angular, and using Eq.~\zero, we write
$$ \eqalign{ W_t &= \sum_{J=q}^\infty \sum_{M=-J}^J
                             a^{JM}(r,t) Y_{qJM} \cr
            W_r &= {1\over r} \sum_{J=q}^\infty \sum_{M=-J}^J
                            b^{JM}(r,t) Y_{qJM} \cr
            W_a &=   \sum_{J=q-1}^\infty \sum_{M=-J}^J
                     f^{JM}_+(r,t) \left[\c^{(1)}_{qJM}\right]_a
                +    \sum_{J=q+1}^\infty \sum_{M=-J}^J
                     f^{JM}_-(r,t) \left[\c^{(-1)}_{qJM}\right]_a , }
\eqno\eq$$
where an index $a$ represents either $\theta$ or $\phi$.  We now insert
these expansions into the action and keep only terms quadratic in
$W_\mu$.  Using the properties of the harmonics given above
(and keeping track of the
additional metric factors arising in our curved space-time) we obtain
a sum of terms, each corresponding to definite values of $J$ and $M$:
$$ S_{\rm quad}= \sum_{J=q-1}^\infty \sum_{M=-J}^J S_{\rm quad}^{JM} .
\eqno\eq$$

The action for the lowest angular momentum, $J=q-1$, is particularly
simple.  As noted above, there is only a single multiplet of vector
harmonics, and these have the properties of being purely angular and
of having vanishing covariant curl.  One finds that
\def\rint{\int_{r_H}^\infty dr }
\def\rtint{\int dt \rint}
$$    \eqalign{ S_{\rm quad}^{(q-1)M}
        &= \rtint  \left\{ {1\over B} |\dot f_+|^2
     - B |f'_+|^2 - m^2 |f_+|^2 +  {qg\over 2r^2} | f_+|^2
                \right\} . }
\eqn\lowest$$
(Here, and henceforth, we omit the superscripts $JM$ on the
various coefficient functions.)

The general case is much more complicated, with
$$  \eqalign{ S_{\rm quad}^{JM}
         &= \rtint \left\{ | \dot b -ra' |^2
    + {1\over B}\left[\left|\dot f_+ - \splus a \right|^2
          + \left|\dot f_- - \sminus a \right|^2 \right] \right.\cr
   &\qquad
   - B\left[ \left| f'_+ - {1\over r}\splus b\right|^2
     + \left| f'_- -{1\over r} \sminus b\right|^2 \right]
     -{1\over r^2} \left| \splus f_+ - \sminus f_- \right|^2 \cr
  &\qquad \left.
    - m^2 \left[ | f_+|^2 + | f_-|^2
          + B|b|^2 - r^2{1\over B}|a|^2 \right]
     +  {gq\over 2r^2}\left[ | f_+|^2 - | f_-|^2 \right]
                \right\} . }
\eqn\quadpertaction$$
This simplifies a bit when $J=q$, since the terms containing $f_-$ are
then absent.  As one would expect, it reduces to Eq.~\lowest\ if $J$ is
set equal to $q-1$ and $a$, $b$, and $f_-$ are set equal to zero.

Our task is now to analyze the behavior of perturbations governed by the
various $S^{JM}_{\rm quad}$.  As a first step, we will now show that
the functions $a$ and $b$  (corresponding to $W_t$ and $W_r$,
respectively) can be eliminated from the analysis.

\chapter{Elimination of $W_t$ and $W_r$}

     The time derivatives of $W_t$ do not enter the action \quadpertaction.
Hence, it (or $a(r,t)$ in the reduced action \quadpertaction) is a
nondynamical constrained field that can be eliminated.  We
proceed as follows.  If the dynamical fields $f_+$, $f_-$,
and $b$ are assembled into a vector
$$   z = \left({f_+\over\sqrt{B}},\, {f_-\over\sqrt{B}},\, b\right)^T ,
\eqno\eq$$
the action \quadpertaction\ may be written in the form
$$  S_{\rm quad}^{JM}  = \rtint \left[{\dot z}^\dagger  \dot z
   + ({\dot z}^\dagger F a + a^* F^\dagger \dot z)
   + a^* G a  - z^\dagger H z \right] ,
\eqn\vectoraction$$
where
$$ F = \left( -{k_+\over \sqrt{B} }, \,-{k_-\over \sqrt{B} },\,
           r {\partial\over \partial r} \right)^T ,
\eqno \eq$$
$$ \eqalign {G &= -{\partial\over \partial r} r^2 {\partial\over \partial r}
     + {1\over B} \left( k_+^2 + k_-^2 + r^2m^2 \right) \cr
         &= \vphantom{\biggl(} F^\dagger F + B^{-1} r^2m^2 }
\eqno\eq$$
and $H$ is a $3 \times 3$ matrix, the explicit form of which we do not need
at the moment.  (In going from Eq.~\quadpertaction\ to Eq.~\vectoraction\
we have performed an integration by parts and dropped the surface
terms.  In the next section we will determine the boundary conditions
on the fields; with these boundary conditions, the neglect of the
surface terms is justified.)

Variations with respect to
$a^*(r,t)$ and $z^\dagger(r,t)$ yield the field equations
$$ 0= F^\dagger \dot z + G a
\eqno\eq $$
and
$$ 0= \ddot z + F \dot a + Hz .
\eqno\eq $$
Using the first of these to solve for $a$, and then
substituting the result into the second equation, we obtain
$$   0= \left( I - F G^{-1} F^\dagger \right) \ddot z + Hz .
\eqn\fieldequation$$

Instabilities of the Reissner-Nordstr\"om solution correspond to
solutions of Eq.~\fieldequation\ that have a time dependence of the
form $e^{\alpha t}$ with real $\alpha >0$.  If the operator acting on
$\ddot z$ is positive definite, these solutions will be in one-to-one
correspondence with the negative eigenvalues of $H$.
To show that this operator is positive,
we define the projection operator
$$ P = F (F^\dagger F)^{-1} F^\dagger
\eqno\eq$$
and write the operator in question in the manifestly positive form
$$  I - F G^{-1} F^\dagger  =
   (I-P) + PF \left[ {1\over F^\dagger F} -
   {1\over F^\dagger F + B^{-1} r^2m^2 } \right]F^\dagger P .
\eqno\eq $$
(Note that, although the unstable modes correspond to the negative
eigenvalues of $H$, the actual value of $\alpha$
is not given by the corresponding eigenvalue unless the unstable mode
lies in the subspace spanned by $I-P$.  This will not affect our
calculations, because we are addressing
only the question of the existence of instabilities.)

Our problem is thus reduced to the study of the positivity of the
potential energy
$$ \tilde E_{\rm quad}^{JM} = \rint z^\dagger H z ,
\eqno\eq $$
which is independent of $W_t$.  Because the radial derivative
of $b$ does not enter the action, a further
reduction is possible.  We write $\tilde E_{\rm quad}^{JM}$ as
the sum of two
integrals, one of which contains only $f_+$ and $f_-$, and the other of
which is a perfect square:
$$\eqalign{ \tilde E_{\rm quad}^{JM}
        &= \rint \left\{
     B \left[ |f'_+|^2 + |f'_-|^2
     - {1\over r^2m^2+\J^2} |k_+ f'_+ + k_- f'_-|^2  \right] \right.
\cr
  &\qquad\qquad + \left.{1\over r^2} \left| \splus f_+ - \sminus f_- \right|^2
    + m^2 \left[ | f_+|^2 + | f_-|^2 \right]
   +  {qg\over 2r^2}\left[ | f_+|^2 - | f_-|^2 \right] \right\} \cr
    &+ \rint  \, B \left| k_+ f'_+ + k_- f'_-
      - {(r^2m^2 + \J^2 )\over r}\, b \right|^2 . }
\eqno\eq $$
The second integral on the right is clearly nonnegative.  Furthermore,
given any  $f_+(r)$ and $f_-(r)$, it is always possible to choose $b$ so
that this integral vanishes.  Hence, a necessary and sufficient
condition for instability is that
there be configurations of $f_+(r)$ and $f_-(r)$ for which the first
integral is negative.  We denote this quantity by $E_{\rm quad}^{JM}$,
and write
$$  E_{\rm quad}^{JM} = \rint
    \left[ B\,{f'}^\dagger K_J f'
     + f^\dagger\left( m^2 I - {V_J \over r^2} \right) f \right] ,
\eqn\efunc $$
where $f \equiv (f_+, f_-)^T$ and we have defined matrices
$$  K_J = I - {1 \over r^2m^2 +\J^2}
      \left( \matrix{ \high k_+^2  & k_+k_- \cr
                 \high    k_+k_- & k_-^2 } \right),
    \qquad\qquad  J>q
\eqno\eq$$
and
$$ V_J =  \left(  \matrix{\high -k_+^2 + {qg\over 2}  &  k_+k_- \cr
                           k_+k_- & \high -k_-^2 - {qg\over 2} } \right),
      \qquad\qquad  J>q .
      \eqno\eq $$

     For $J=q-1$ or $q$, matters are simplified by the absence of
$f_-$, so that $K_J$ and $V_J$ are numbers rather than matrices.
Using the fact that $k_+^2 $ is equal to $0$ and $q$, respectively,
for these two cases, we obtain
$$   K_{q-1}= 1,  \qquad\qquad
     V_{q-1} = {gq\over 2}
   \eqno\eq$$
and
$$   K_{q}= {r^2 m^2 \over r^2m^2 + q},  \qquad\qquad
     V_{q} = {(g-2)q\over 2} .
   \eqno\eq$$
(The results for $J=q-1$ could, of course, be obtained directly from
Eq.~\lowest.)

    The positivity of $K_J$ will be of importance later.
This property is obvious for $J=q-1$ and $J=q$, while for $J>q$ it
follows from the positivity of the two eigenvalues
$$   k_1 =1 \qquad {\rm and }\qquad
        k_2= {r^2m^2 \over r^2m^2 + \J^2 } .
   \eqn\keval $$

\chapter{Existence of instabilities}

Our task is now to analyze the positivity of the potential energies
$ E_{\rm quad}^{JM}$.  In doing this, it convenient to transform  to
a tortoise coordinate $x$ defined by
$$  {dx\over dr} = B^{-1}(r) .
  \eqno\eq $$
Spatial infinity, $r=\infty$, corresponds to $x=\infty$, while the
horizon radius $r_H$ corresponds to $x=- \infty$.  In terms of this
coordinate
$$  E_{\rm quad}^{JM} = \int_{-\infty}^\infty dx
    \left[  {df^\dagger\over dx}  K_J {df\over dx}
     + f^\dagger U_{eff}   f \right] ,
\eqn\equad $$
where
$$ U_{eff} =B(r)\left[ m^2 I - {V_J \over r^2} \right] .
\eqno\eq$$
Instabilities correspond to the existence of functions $f(x)$ which
for which the energy functional is negative.  This is equivalent to the
existence of a negative energy bound state in the spectrum of the
Hamiltonian
$$  H = -{d\over dx} \left(K_J {d \over dx} \right)  + U_{eff}(r(x)) .
   \eqn\ham $$

    Because $K$ is a positive matrix, a necessary condition for the
occurrence of an instability is that $U_{eff}$ be negative (or have a
negative eigenvalue) in some region outside the horizon.  This
requires that $V_J$ have a positive eigenvalue $a^2$ and that the
horizon radius $r_H$ be less than a value $r_0(J)= a/m$. For $J=q-1$,
this gives the requirements that $g$ be positive and that $r_H$ be
less than
$$   r_0(q-1) = \sqrt{gq \over 2} m^{-1} .
  \eqn\jqmrz $$
For $J=q$, instability can arise only if $g >2$, while
$$    r_0(q) = \sqrt{(g-2)q \over 2} m^{-1} .
  \eqn\jqrz $$
If $J>q$   the two eigenvalues of $V_J$ are
$$  v_\pm = {1\over 2}
      \left[-\J^2 \pm \sqrt{\J^4 + g(g-2)q^2} \right] .
\eqn\vpm $$
If $0\le g\le 2$, both of these eigenvalues are negative
and no instabilities can arise.  For values of $g$ outside this range,
$v_+$ is positive and
$$   r_0(J) =
    {1\over 2} \left[-\J^2 + \sqrt{\J^4 + g(g-2)q^2} \right]^{1/2} m^{-1},
\qquad   J>q  .
 \eqno\eq $$
Note that for fixed $q$ and $g$, $r_0(J)$ is a decreasing function of
$J$.

Because of the gradient term in the energy, these conditions by
themselves are not sufficient for the existence of an unstable mode.
In fact, the horizon radius must be less than a critical value
$r_{\rm cr}(J)$ that is somewhat smaller than $ r_0(J)$.  In the next section
we will describe the numerical determination of $r_{\rm cr}$.  Before
doing so, we show that an unstable mode will always arise if
$r_H$ is sufficiently small, i.e., that there is in fact a positive
$r_{\rm cr}$. (Our proof implicitly assumes that $r_H$ is greater than
the extremal horizon radius.  This is a significant constraint on
the existence of $r_{\rm cr}$ only if $m$ is comparable to $M_{\rm Pl}$.)
To see that
this should not be immediately obvious, consider the flat space
analogue of our problem in which $B(r)$ and $K_J(r)$ are both
identically equal to unity, and the integration is restricted to a
region $r >r_* \ge 0$ with the boundary condition that $f(r_*) =
f(\infty)=0$. From the identity
$$  \int^\infty_{r_*} dr |f'|^2
   = \int^\infty_{r_*} dr \left[ \left| f' -{f \over 2r} \right|^2
        + {|f|^2\over 4r^2} \right] ,
\eqno\eq $$
it is easy to see that the energy functional is positive, no matter
how small $r_*$ is, if $V$ (or its smallest eigenvalue) is less than
$1/4$.

      The situation is different in curved space, essentially because
the metric factor suppresses the effects of the gradient terms near
the horizon.  To demonstrate this, let us consider the functional
$$ E' = \rint \left[ {\tilde B} |f'|^2
    + \left(m^2 - {V\over r^2} \right) |f|^2 \right] ,
       \eqno\eq$$
with $V >0$ being equal to the positive eigenvalue of $V_J$ and
$$ {\tilde B} = 1 - {r_H \over r}  .
    \eqno\eq $$
For a given value of $r_H$, the Reissner-Nordstr\"om $B(r)$ is less than
$\tilde B(r)$ everywhere outside the horizon.  From this, together
with the fact that $K_J(r) \le 1$, we see that if the potential energy
functional \equad\ is positive, then so must be $E'$.  Conversely, if the
Hamiltonian $H'$ that we obtain from $E'$ has a bound state, then so
must the $H$ of Eq.~\ham, and hence there will be an instability.
Our aim is to show that such a bound state develops if $r_H$ is made
sufficiently small (but still positive), with $m$ and $V$ held fixed.
Dimensional analysis shows that this limit is equivalent to holding
$V$ and $r_H$ fixed and letting $m$ become arbitrarily small.  We will
find it more convenient to use this latter method.

     Defining a tortoise coordinate $y$ by $dy/dr = \tilde
B^{-1}(r)$, we have
$$  H' = -{d^2\over dy^2}  + U(r(y))  ,
   \eqno\eq $$

with
$$  U = \tilde B \left( m^2 - {V\over r^2} \right) .
     \eqno\eq $$
Now write
$$  U = U_1 + U_2 + U_3 ,
     \eqno\eq $$
where
$$  U_1(y) = \cases{ U(y) , & $r_H < r < Nr_H $ \cr
                     0    , & $r> Nr_H $ }
    \eqno\eq  $$
$$  U_2(y) = \cases{ U(y) , & $N r_H < r < r_0 $ \cr
                     0    , & otherwise }
    \eqno\eq  $$
$$  U_3(y) = \cases{ 0 , & $r < r_0  $ \cr
                     U(y)    , & $r> r_0 $, }
    \eqno\eq  $$
with $N$ being some fixed, large number and $r_0 = \sqrt{V} /m $.  It
is assumed that $m$ is small enough that $r_0 \gg Nr_H \gg r_H$.

    Consider the Hamiltonian $H'_0$ obtained by replacing $U(y)$ by
$U_1(y)$.  Because $U_1(y)$ is everywhere negative on the real line, a
standard result of elementary quantum mechanics guarantees the
existence of at least one negative energy bound state.  Let us denote
the energy of this state by $E_B = - \alpha^2$, and let $\psi(y)$ be
the corresponding normalized wave function.  We do not need a detailed
expression for $E_B$, but only the fact that it has a smooth nonzero
limit as $m \rightarrow 0$.  This can be seen by noting that setting
$m=0$ changes the value of $U_1(y)$ at any point by a fractional
amount of at most $m^2 (Nr_h)^2/V = (Nr_h/r_0)^2 \ll 1$.  For a
sufficiently large value of $r_0$, the range $r>r_0$ will be entirely
in the classically forbidden region. Hence, in this region, $\psi$
will be of the form $ \psi(y)=A e^{-\alpha y}$ with $A$ only weakly
dependent on $m$.

     An upper bound on the minimum eigenvalue $E_{\rm min}$ of $H'$
can now be obtained by using $\psi$ as a trial
wavefunction:
$$ \eqalign{  E_{\rm min}
    &\le E_B + \int_{- \infty}^\infty dy ( U_2 + U_3)|\psi|^2 \cr
    & < -\alpha^2 + \int_{y(r_0)}^\infty dy  U_3 |\psi|^2 \cr
    & < -\alpha^2 + m^2 \int_{y(r_0)}^\infty dy  |A|^2 e^{-2\alpha y}\cr
    & = -\alpha^2 + {m^2 \over 2 \alpha} e^{-2 \alpha y(r_0)} . }
     \eqno\eq $$
For sufficiently small $m$ (i.e., large $r_0$ and $y(r_0)$), the
second term is smaller in magnitude than the first. Hence $E_{\rm
min}$ is negative and we have proven our result.

\chapter{Numerical Results}

In the original coordinates,
we have an energy functional \efunc\ for the perturbations.
As we have shown, the black hole is unstable if and only if there
exist functions $f_+$ and $f_-$ such that the appropriate functional
is negative.  We find such functions by looking for solutions to the
eigenvalue equation with negative eigenvalue
for the Hamiltonian defined by the functional.\footnote{1}{This equation can
also be found by varying the energy functional with respect to $f$ and
inserting a Lagrange multiplier to implement the normalization
constraint.}
The critical radius $r_{\rm cr}$ is the value of the horizon
radius above which there are no negative eigenvalues and below which
there is at least one.  When inspecting the equations, it is
simpler to consider the case where $r_H$, and thus the boundary
condition, is fixed and $m$ varies. (The relevant
dimensionless parameter is $m r_H$.)  It is then clear that
decreasing $m$ strictly
and smoothly decreases the energy, and thus the energy eigenvalues;
the critical value of $m r_H$ will be where the lowest eigenvalue
crosses from positive to negative.

A critical point in the parameter space, then, will be one for which
the Hamiltonian has
(generically) one zero eigenvalue and no negative eigenvalues.
In general, the wavefunction has two components, $f_+$ and $f_-$.
For $J \le q$, however, there is only one component and so to find
the critical point
we search for values of the parameters such that the
solution to the eigenvalue equation with eigenvalue zero has the
following properties:

1) The solution exponentially decays to zero at $r \to \infty$.  We can
solve the equation analytically for large r; we choose the solution
that is decaying.

2) The solution goes to a finite value at $r \to r_H$.  This is actually a
fairly restrictive condition; a typical solution blows up at the horizon.  That
this is the correct boundary condition can be seen by looking in
tortoise coordinates:  In these coordinates a solution with a negative
energy eigenvalue decays exponentially as $x \to -\infty$.  In the limit
that the negative eigenvalue goes to zero, this exponential decay is
arbitrarily slow.

3) The solution does not change sign.

Properties (1) and (2) are the appropriate boundary conditions; property
(3) ensures that there are no negative eigenvalues. (In one dimension,
the ground state wavefunction is the unique energy eigenfunction with no
zeroes.)

{}From the argument in the previous section, we see that there will always
be a bound state for sufficiently small $r_H$, or equivalently,
sufficiently small $m$.  Thus we can fix the other parameters and
search for the critical value of $m$ for which there is a solution to
the differential equation that satisfies (1) through (3). We
enforce condition (1) by integrating inward from $r \gg r_0$, choosing
initial values to approximate the decaying exponential we expect at
large $r$. (The precise initial values chosen are not important, since
the decaying exponential solution will quickly dominate as we integrate
to smaller $r$.)  If the solution changes sign before we reach $r_H$, we
know that $m$ is below the critical value; if the solution
diverges as $r \to r_H$ without changing sign, we know that $m$ is above
the critical value.\footnote{2}{The algorithm we used
in fact searches for a sign
change, but if it does not find one, it tests the value of $d f / d \ln
(r - r_H)$ very close to $r_H$.  If this is positive, it is assumed that
the solution will in fact change sign closer to the horizon than the
integrator can proceed, and thus $m$ is still below the critical
value.}
We can use this information to quite rapidly and
accurately zero in on the critical value of $m$ (or, equivalently,
$r_H$).

For $J = q-1$, the energy functional is given by
$$ E_{\rm quad}^{(q-1)M} =  \rint  \left[ B\,| f_+'|^2
     +\left( m^2  -{qg \over 2r^2} \right) |f_+|^2 \right] ,
\eqno\eq $$
thus the (zero eigenvalue) equation to be solved is
$$
 - {d \over dr} (B f'_+) + \left( m^2  -  {qg\over 2r^2} \right)  f_+ = 0 .
\eqno\eq $$
In this case
there is only one other dimensionless parameter, $gq$.
(We make the approximation $r_- = 0$, good for $m \ll M_{\rm Pl}$.)
We have calculated the
critical mass for a variety of values of this parameter.

For $J = q$, we have
$$ E_{\rm quad}^{qM} = \rint \left[
    B \left({r^2 m^2 \over r^2m^2 + q} \right)\,| f_+'|^2
     + \left( m^2 - {(g-2)q \over 2 r^2} \right) |f_+|^2 \right] ,
\eqno\eq $$
which gives the zero eigenvalue equation
$$ -{d \over dr}
    \left[ B \left({r^2 m^2 \over r^2m^2 + q} \right)\, f_+' \right]
     + \left( m^2 - {(g-2)q \over 2 r^2} \right) f_+ = 0 .
\eqno\eq $$
In this equation $q$ appears separately from $g$; in principle, the
critical mass depends on both.

In the case $J > q$, the wavefunction has two components.
This gives a two-component equation,
which does not have a simple relation between nodes and ground states
and which would require tuning of the
relative weights of the two components, as well as the mass.
In this case, we are able to
calculate bounds on the energy by minimizing one-component functionals;
these energy bounds give bounds on the
critical parameter.

The energy functional for $J>q$ is given by Eq.~\efunc.
Consider a basis
in which $V_J$ is diagonal, and where the upper left element is
$v_+$, the only positive eigenvalue of $V_J$. (See Eq.~\vpm)
The matrix $K_J$ has two positive eigenvalues, $k_1$ and $k_2$
(Eq.~\keval); $k_1$ is the larger of the two eigenvalues.

Now, we are searching for the minimum energy configuration for the
functional $E[f]$ (subject to some normalization constraint).  We call
this configuration $f_0$.  Now let $f_1$ be a configuration which
minimizes the functional, subject to the additional constraint that
the lower component of $f$ is zero (in this particular
basis).\footnote{3}{Since the upper left
component of $V_J$ is the only one which is positive, the energy will
never be negative if we require that the upper component of $f$ is
zero.}  Clearly $E[f_1] \ge E[f_0]$.
Now consider the modified functional $E_-[f]$ obtained by substituting
the smaller eigenvalue of $K_J$, $k_2$ (actually $k_2 I$), for $K_J$.
Clearly $E_-[f] \le
E[f]$ for all $f$, so if we minimize $E_-$ with a configuration $f_2$,
$E_-[f_2] \le E[f_0]$.  Thus we have bounded $E[f_0]$ above and below.
But both $E[f_1]$ and $E_-[f_2]$ can be calculated with a single
component functional, which allows us to put bounds on
the critical parameters without tuning the two-component equation.

  Now if we write
$$ K_J = I - {1\over r^2 m^2 + \J^2} C ,
\eqno\eq $$
then in the basis where $V_J$ is diagonal (with the positive
eigenvalue $v_+$ at the upper left),
$$ C = \left( \matrix{\high {\J^2 \over 2} + {1\over 4 v}[\J^4 + q^2(g-2)] &
 \high {1\over 4 v}q (g-2) \sqrt{\J^4 - q^2} \cr
 \high {1\over 4 v}q (g-2) \sqrt{\J^4 - q^2} &
\high {\J^2 \over 2} - {1\over 4 v}[\J^4 + q^2(g-2)] \cr } \right) ,
\eqno\eq $$
where $v = \sqrt{\J^4 + q^2 g (g-2)}$.
So, to calculate the minimum energy with the lower component constrained
to be zero, $E[f_1]$, we minimize the functional
$$ E_{\rm quad}^{JM}[f_u] = \rint \left[
    B \left(1 - {1\over r^2 m^2 + \J^2} C_{uu} \right)\,| f_u'|^2
     + \left( m^2 - {v_+ \over  r^2} \right) |f_u|^2 \right] ,
\eqno\eq $$
where $C_{uu} = {\J^2 \over 2} +{1\over 4 v}[\J^4 + q^2(g-2)]$
is the upper left component of the matrix $C$ defined
above.  If we tune $m$ to solve the zero eigenvalue equation
associated with this functional, we obtain a {\it lower} bound on the
critical mass.
We bound the energy from below (and thus the critical mass from above)
with the $E_-$ described above.  Since both
$V_J$ and $K_J$ are diagonal in this functional, the upper and lower
components of $f$ completely decouple: $E_- = E_-[f_u] + E_-[f_l]$,
where $f_u$, $f_l$ are the upper and lower components of $f$
respectively.  Because $E_-[f_l]$ is positive definite,
we can set $f_l = 0$ and just use the functional
$$ E_{-,\rm quad}^{JM}[f_u] = \rint  \left[
    B \left( {r^2 m^2 \over r^2 m^2 + \J^2}
               \right)\,| f_u'|^2
     + \left( m^2 - {v_+ \over  r^2} \right) |f_u|^2 \right] .
\eqno\eq $$
Solving the zero eigenvalue equation associated with this functional
gives an {\it upper} bound on the critical mass.

We have calculated and graphed
some critical values (or bounds thereon) of $m r_H$, where $m$ is the
$W$ mass and $r_H$ is the outer horizon radius, for varying $g$, $q$,
and $J$.  (We have calculated
these assuming that $r_- = 0$, i.e.\ a Schwarzschild solution, which
will be approximately correct as long as $m$ is far below the Planck
scale.  Provided that the black hole is not very near extremality, the
calculated values will not be significantly affected by a non-zero $r_-$.
Handling the extremal case is considerably more delicate,
and we do not discuss it here.)

First note that the upper and lower bounds on $r_{\rm cr}$ calculated
for $J > q$ are quite close, which justifies calculating only these bounds
rather than doing the full, two-component calculation.  All the curves
show the qualitative dependence of the critical radius on $g$ (and $J$).
Most importantly, it appears from our numerical results
that $r_{\rm cr}$ is a decreasing
function of $J$; recall that $r_0(J)$ has the same property.
Thus, as we reduce the radius of
a black hole with fixed charge, it will be the mode with the
lowest angular momentum that first becomes unstable. If $g$ is positive
and $q > {1\over 2}$,
then this will always be the mode with $J = q-1$.
If $g$ is
negative the $J \le q$ modes never become unstable (see Eqs.
\jqmrz\ and \jqrz) and it is
the $J = q + 1$ mode that first becomes
unstable. (We do not display results for $g < 0$.  For the $J > q$
modes, making the substitution $g \to 2 - g$ does not change the
calculated upper bound on $r_{\rm cr}$, although it does change the
lower bound.  The qualitative behavior is quite similar.)
Finally, if $q = {1\over 2}$, then there is no instability if
$0 \le g \le 2$; for $g > 2$, it is the $J = q = {1\over 2}$ mode that
is the first to become unstable.

\FIG\fone{The critical value of $m r_H$ as a function of $g$ for
$q=1$ and $J=0$, 1, 2, and 5.  For $J=2$ and 5, upper and lower bounds
are plotted for $m r_{\rm cr}$, rather than the actual values.}
\FIG\ftwo{The ratio of $r_{\rm cr}$ to $r_0$ as a
function of $g$ for the values of $J$ shown in Figure~1.  As in Figure~1,
bounds are plotted for $J > q = 1$. Note that the curve for $J=1$ is
largely obscured by the curve for $J=0$ (for $g > 2$).}
It is useful to compare the behavior if $r_{\rm cr}$ with that
of $r_0$.  In Figure \ftwo\ we plot the ratio of these quantities.
Note that for $J \ge q$, we find that $r_{\rm cr} \sim r_0$ for all
values of $g$, with the ratio varying from about $1\over 2$ to 1.
For $J=q  -1$ we find similar behavior for large $g$, but for very small
$g$, we find that the relationship between $g$ and $r_{\rm cr}$
becomes linear, i.e.\ that  $r_{\rm cr} \propto r_0^2$. (Indeed,
for small $g$, we have $m r_{\rm cr} = (m r_0)^2$, with unit
proportionality.)  We have no analytical explanation for this
behavior.  Note also that the values of $r_{\rm cr}/r_0$ for
$J = 1$ are very close to the values for $J=0$, when $g > 2$.
Although the
differences between these curves are very small, they are much greater
than the estimated errors in our calculation.

\chapter{Conclusion}

\baselineskip 20pt plus 1pt minus 1pt
We have investigated the linear stability of magnetically charged black
holes in a theory with a massive charged vector field.  The \rn
solution is stable against small perturbations of the electromagnetic,
gravitational, and scalar fields, so linear perturbations of only the
$W$ field need be considered.  We expanded the $W$ field in angular
momentum eigenstates, the monopole vector harmonics.
For each value of $J$, we found the critical horizon radius at
which an unstable mode appears.  This radius (in units of the $W$
mass) depends upon the angular momentum of the mode, $J$, the charge of
the black hole, $q$, and the magnetic moment of the $W$ field, $g$.
This makes quantitative the arguments of Ref.~[\leew] concerning the
stability of \rn black holes in this theory.

There are some unanswered questions in this analysis.  In particular,
when the critical horizon radius is close to the radius of an extremal
\rn black hole with the same charge, these calculations must be
modified.  Indeed, if the critical radius we have calculated here
is less than
the extremal radius, there may be no instability at all.  Because of the
quite different nature of the coordinates near the extremal horizon,
our algorithm for calculating the critical value was ineffective, and
so this case was not directly addressed.  Another outstanding
question is the origin of the linear behavior of the critical radius
for $J=q-1$ near $g = 0$.

In general, it is the mode with the lowest $J$ that becomes unstable at
the largest $r_H$.  Consider the evaporation of a \rn black hole with
magnetic charge $q/e$.  Once the horizon radius falls below $r_{\rm
cr}$ for the lowest mode, we expect that the black hole will be
described by a nontrivial classically stable solution with given $r_H$
and $q$, rather than the unstable \rn solution.  If $q=1$, then the
lowest mode has $J=0$ and is spherically symmetric, so we expect that
this new stable solution will also be spherically symmetric.  Suppose
instead that $q \ne 1$.  The lowest mode will have $J>0$ and the new
classically stable solution will not be spherically symmetric.  The
further evolution of the black hole depenmds on whether or not the
theory possesses finite energy magnetic monopoles that are not black
holes.  If it does, then as the horizon contracts with further
evaporation we expect that the black hole will decay by emitting
monopoles and settling down to a $q=1$ spherically symmetric solution.
Eventually the horizon will disappear, leaving behind a simple
monopole.  If there are no such monopole solutions, the endpoint of the
decay will be an asymmetric extremal black hole.

\refout
\figout
\vfil\eject
\iffigsinc
    \nopagenumbers
    \vfil
    \centerline{\epsfbox{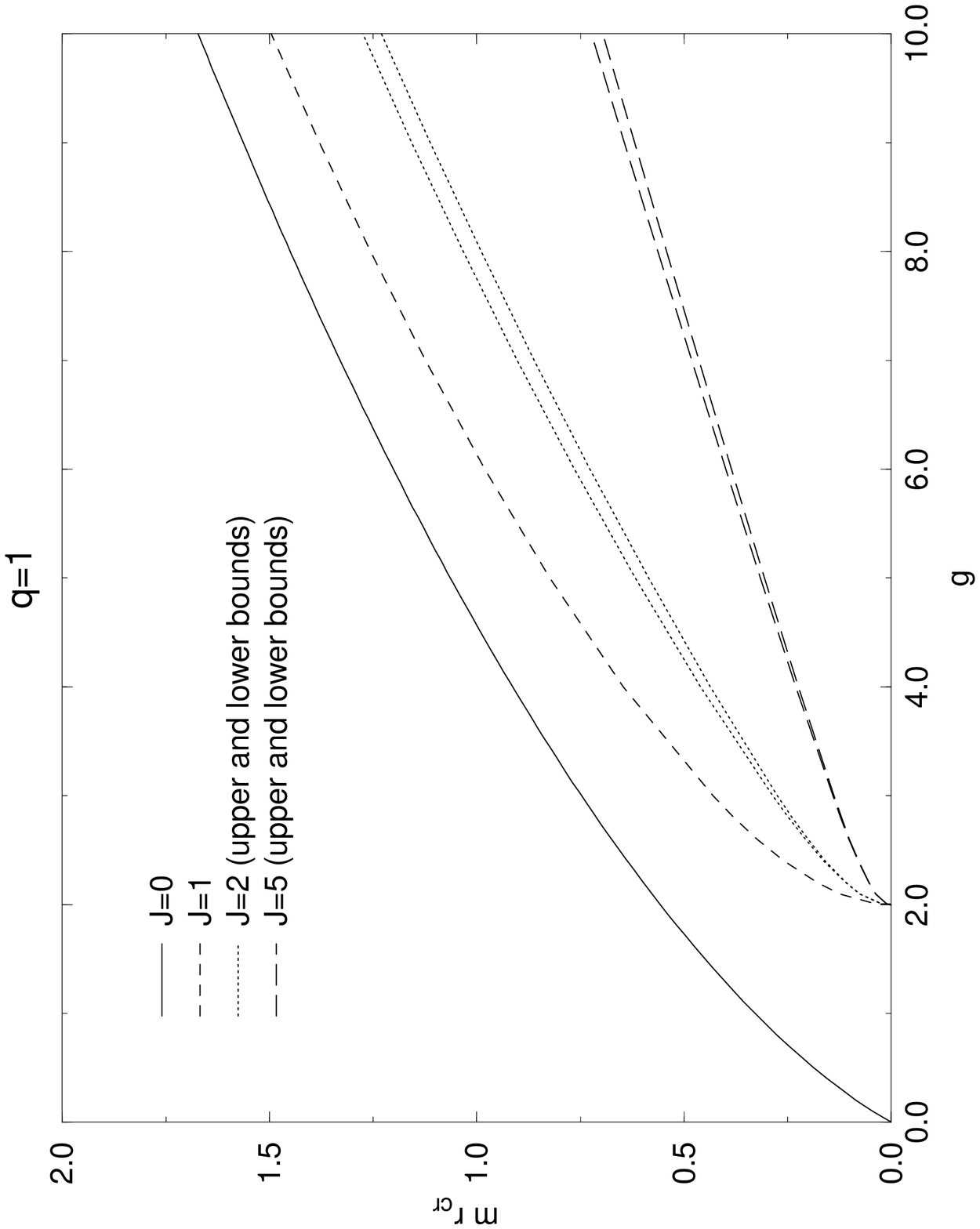}}
    \vfil\eject
    \vfil
    \centerline{\epsfbox{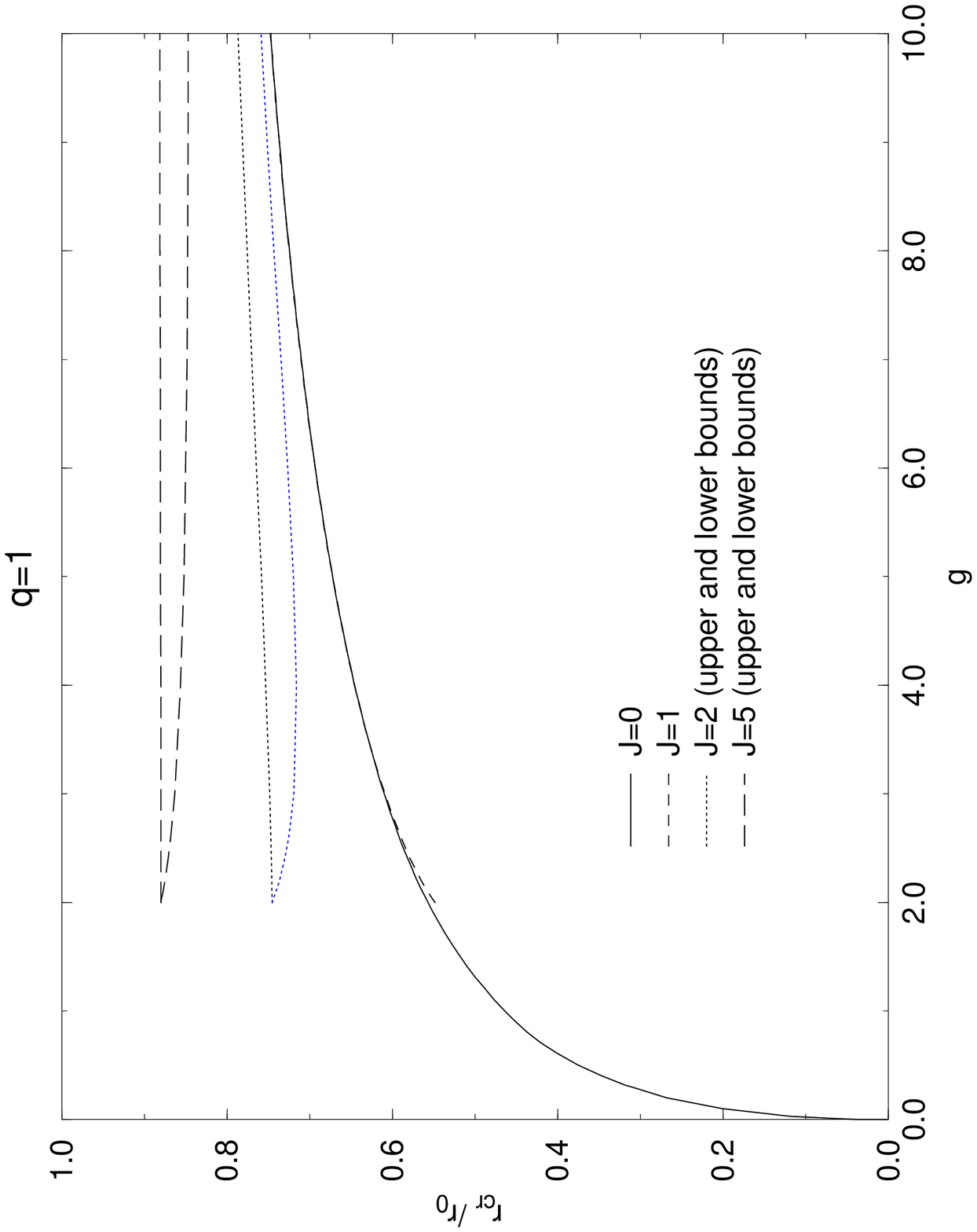}}
    \vfil\eject
\fi
\end